\documentclass{article}
\usepackage{spconf,amsmath}
\usepackage{graphicx}
\usepackage{cite}

\usepackage[caption=false,font=footnotesize]{subfig}

\newcommand{\ignore}[1]{}
\newcommand{\wfig}[1]{Fig.~\ref{fig:#1}}%
\newcommand{\wfigure}[1]{Figure~\ref{fig:#1}}%
\newcommand{\wfigsub}[2]{Fig.~\ref{fig:#1}\subref{fig:#2}}
\newcommand{\wfiguresub}[2]{Figure~\ref{fig:#1}\subref{fig:#2}}%
\newcommand{\wfigsubonly}[2]{\ref{fig:#1}\subref{fig:#2}}

\title{Two-layer Near-lossless HDR Coding \\with Backward Compatibility to JPEG}
\name{Hiroyuki KOBAYASHI$^{\dagger}$ \qquad Osamu WATANABE$^{\ddagger}$ \qquad Hitoshi KIYA$^{\star}$}

\address{$^{\dagger}$ Tokyo Metropolitan College of Industrial Technology\\
    $^{\ddagger}$Dept. of Electronics \& Computer Systems, Takushoku
University.\\
    $^{\star}$Dept. of Computer Science, Tokyo Metropolitan University}
\begin{document}
\ninept
\maketitle
\begin{abstract}
We propose an efficient two-layer near-lossless coding method using an extended histogram packing technique with backward compatibility to the legacy JPEG standard.
The JPEG XT, which is the international standard to compress HDR images, adopts a two-layer coding method for backward compatibility to the legacy JPEG standard.
However, there are two problems with this two-layer coding method.
One is that it does not exhibit better near-lossless performance than other methods for HDR image compression with single-layer structure.
The other problem is that the determining the appropriate values of the coding parameters may be required for each input image to achieve good compression performance of near-lossless compression with the two-layer coding method of the JPEG XT.
To solve these problems, we focus on a histogram-packing technique that takes into account the histogram sparseness of HDR images.
We used zero-skip quantization, which is an extension of the histogram-packing technique proposed for lossless coding, for implementing the proposed near-lossless coding method.
The experimental results indicate that the proposed method exhibits not only a better near-lossless compression performance than that of the two-layer coding method of the JPEG XT, but also there are no issue regarding the combination of parameter values without losing backward compatibility to the JPEG standard.
\end{abstract}
\begin{keywords}
  JPEG XT, HDR, Near-Lossless coding, Backward compatibility to JPEG
\end{keywords}
\section{Introduction}
Image-compression methods designed to provide coded data containing HDR content are highly expected to meet the rapid growth in HDR image applications. Generally, HDR images have a much longer bit depth of pixel values and much wider color gamut\cite{ReinhardBook,8026195,ArtusiBook01,BADC11}.
These characteristics are suitable for many applications, such for cinema, medicine and art.
For such applications, HDR images should often be losslessly or near-losslessly encoded.
In other words, they should be compressed with almost no coding loss.

Several methods have been proposed for compressing HDR images
\cite{Ward:2006:JBH:1185657.1185685,6543051,8081467}
and ISO/IEC JTC 1/SC 29/WG 1 (JPEG) has developed a series of international standards referred to as the JPEG XT\cite{6737677,Artusi2015,7426553,JPEGXT,7535096} for compressing HDR images.
The JPEG XT has been designed to be backward compatible with the legacy JPEG standard\cite{JPEG-1} with two-layer coding.
The ISO/IEC IS 18477-8:2016\cite{JPEGXTpart8}, which is known as JPEG XT Part 8, describes how to decode losslessly or near-losslessly encoded HDR images.
The compression performance of JPEG XT Part 8’s near-lossless coding method is not better than that of other methods for HDR-image compression with single-layer coding.
It is also required to find a combination of parameter values that enable good compression performance.
This combination is dependent on input HDR images.

The sparseness of an image histogram has been used for efficient lossless compression\cite{KIYAJan2018,6288143,958146,1034993,988715,1040040,ELCVIA116,eusipco2017,6411962,6213328,6637869,8351220,8456254}.
A `Sparse' histogram means that not all the bins are used.
It is well known that a histogram of an HDR image tends to be sparse\cite{6411962,6637869}.
Two-layer lossless coding methods of HDR images with backward compatibility to the legacy JPEG standard have been proposed to overcome the problems with the JPEG XT Part 8 method\cite{8351220,8456254}.
However, these two-layer methods can not be applied to near-lossless coding due to a limitation of the histogram-packing technique.

We therefore propose an efficient two-layer near-lossless coding method using an extended histogram-packing technique with backward compatibility to the legacy JPEG standard.
To avoid the limitation of the histogram packing technique, we used zero-skip quantization (ZS.Q), which is an extension of the histogram packing technique proposed for lossless compression, for achieving efficient near-lossless coding.
ZS.Q is a technique for one-layer coding\cite{Minewaki2019}, so it has never been applied to two-layer near-lossless coding.
Therefore, no two-layer near-lossless coding using histogram packing techniques has been compared with the JPEG XT Part 8 method.
This paper has two aims.
One is to improve the compression performance of two-layer near-lossless HDR coding without losing backward compatibility with JPEG standard.
The other is to solve the issue regarding the combination of parameter values that the JPEG XT Part 8 method has.
Conventional studies on sparseness of an image histogram focus on improving compression performances.
In other words, they have never discussed the issue regarding the combination of parameter values.

We conducted an experiment to evaluate the proposed method.
The proposed method exhibited not only better near-lossless compression performance than that of the JPEG XT Part 8 method, but also had no issues regarding the combination of parameter values for good compression performance without losing backward compatibility to the legacy JPEG standard.

\section{Near-lossless coding with JPEG XT Part 8 method\label{sec2}}
\begin{figure}[tb]
 \centering
 \includegraphics[width=1\columnwidth]{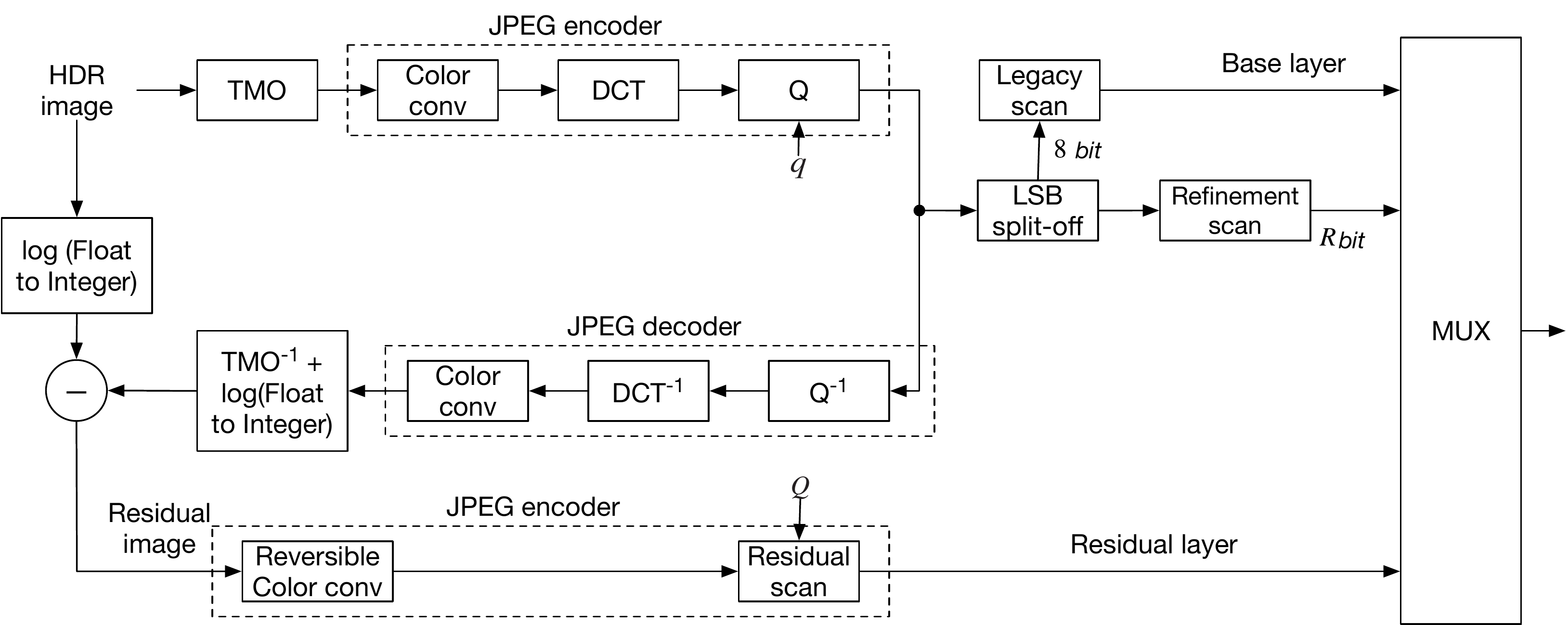}
 \caption{Block diagram of JPEG XT Part 8 method: `TMO' means tone
 mapping operator, Q and Q$^{-1}$ are quantization and inverse
 quantization, respectively, and $q$ and $Q$ are parameters to control quality of
 decoded base layer (LDR) image and residual layer image, respectively.}
 \label{fig:XT_C_enc}
\end{figure}

We first give a brief overview of the JPEG XT Part 8 method then present the problems with near-lossless coding.

\subsection{JPEG XT Part 8 method}
The block diagram of the JPEG Part 8 method is shown in \wfig{XT_C_enc}. 
JPEG XT Part 8 was designed to encode intermediate or high dynamic image sample values without loss or with a precisely controllable bounded loss (near-lossless) using the tools defined in ISO/IEC 18477-1 and minimal extensions of these tools.
For near-lossless compression of HDR images, coding disables the DCT in the residual domain\cite{JPEGXTpart8}.

As shown in \wfig{XT_C_enc}, JPEG XT Part 8 method has three parameters, i.e. $q$, $R$, and $Q$, to control the quality of images.
The $q$ controls the decoded-image quality of the base layer, where the decoded images corresponded to low dynamic range (LDR) ones.
A higher $q$ results in better image quality.
The $R$ is the number of bits used for refinement scan.
Refinement scan is used to improve the precision of coefficients up to 12 bits.
Thus the valid range of $R$ is from $0$ to $4$.
The $Q$ controls the decoded-image quality of the residual layer.

 \subsection{Problems with near-lossless coding}
For near-lossless compression of HDR images, it is required to determine the above three parameters.
To achieve high near-lossless compression, they should be carefully determined.
As mentioned above, $q$ is a parameter to control the quality of the base layer, and
$Q$ and $R$ are two parameters to control the quality of the residual layer.
The first problem is how $Q$ and $R$ are determined to obtain the best performance.

\begin{figure}[tb]
 \centering
 \subfloat[McKeesPub\label{fig:JPEGXTMcKeesPub}]{
  \includegraphics[width=0.5\linewidth]{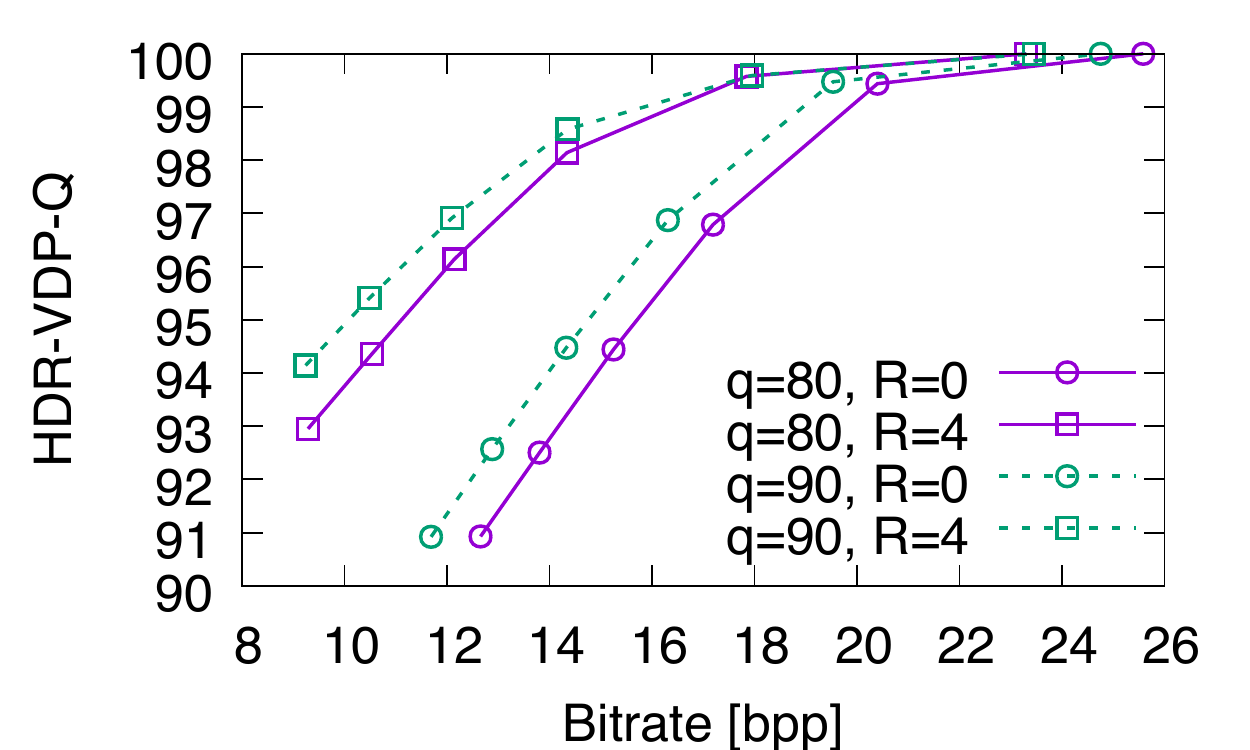}
  }
 \subfloat[memorial\label{fig:JPEGXTmemorial}]{
  \includegraphics[width=0.5\linewidth]{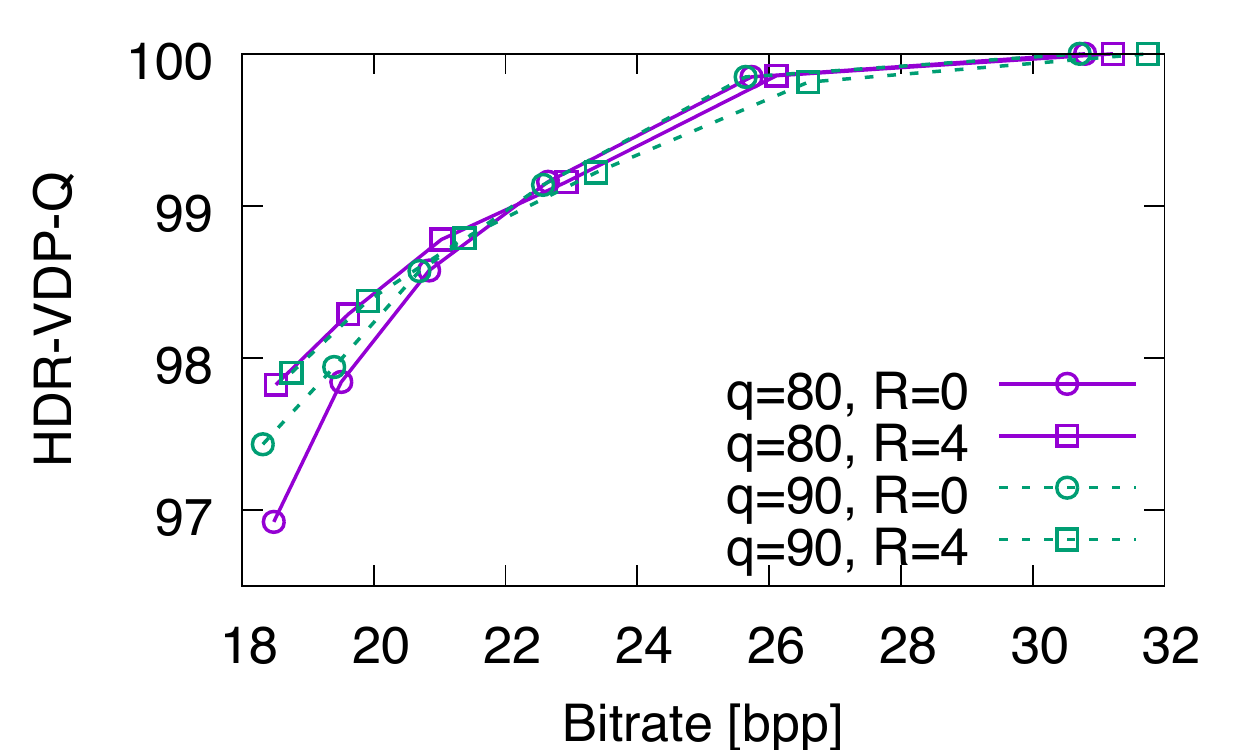}
  }
 \caption{Rate-distortion curves using HDR-VDP-Q of near-lossless compressed HDR images (McKeesPub and memorial) by using JPEG  XT Part 8 method}
 \label{fig:nearLossless}
\end{figure}

\wfigure{nearLossless} illustrates rate-distortion curves of the near-lossless compressed HDR images `McKeesPub' and `memorial' with  $Q=95$ to 100, $q=80, 90$, and $R=0, 4$ in terms of a typical quality metrics of HDR images, HDR-VDP-Q\cite{HDR-VDP2.2}.
From the figure, the optimal combination of three parameters: $q$, $Q$ and $R$, are shown to be image-dependent.
This is the first problem.
The second problem is how compression performance can be improved.

To solve these problems,  the proposed method uses one parameter $\epsilon$ to control the quality of the residual layer instead of two parameters $Q$ and $R$.
Moreover, in an experiment we conducted the proposed method outperformed the JPEG XT Part 8 method in terms of HDR-VDP-Q.

\section{Proposed method}

\subsection{Outline of proposed method}
\begin{figure}[tb]
 \centering
 \includegraphics[width=1\columnwidth]{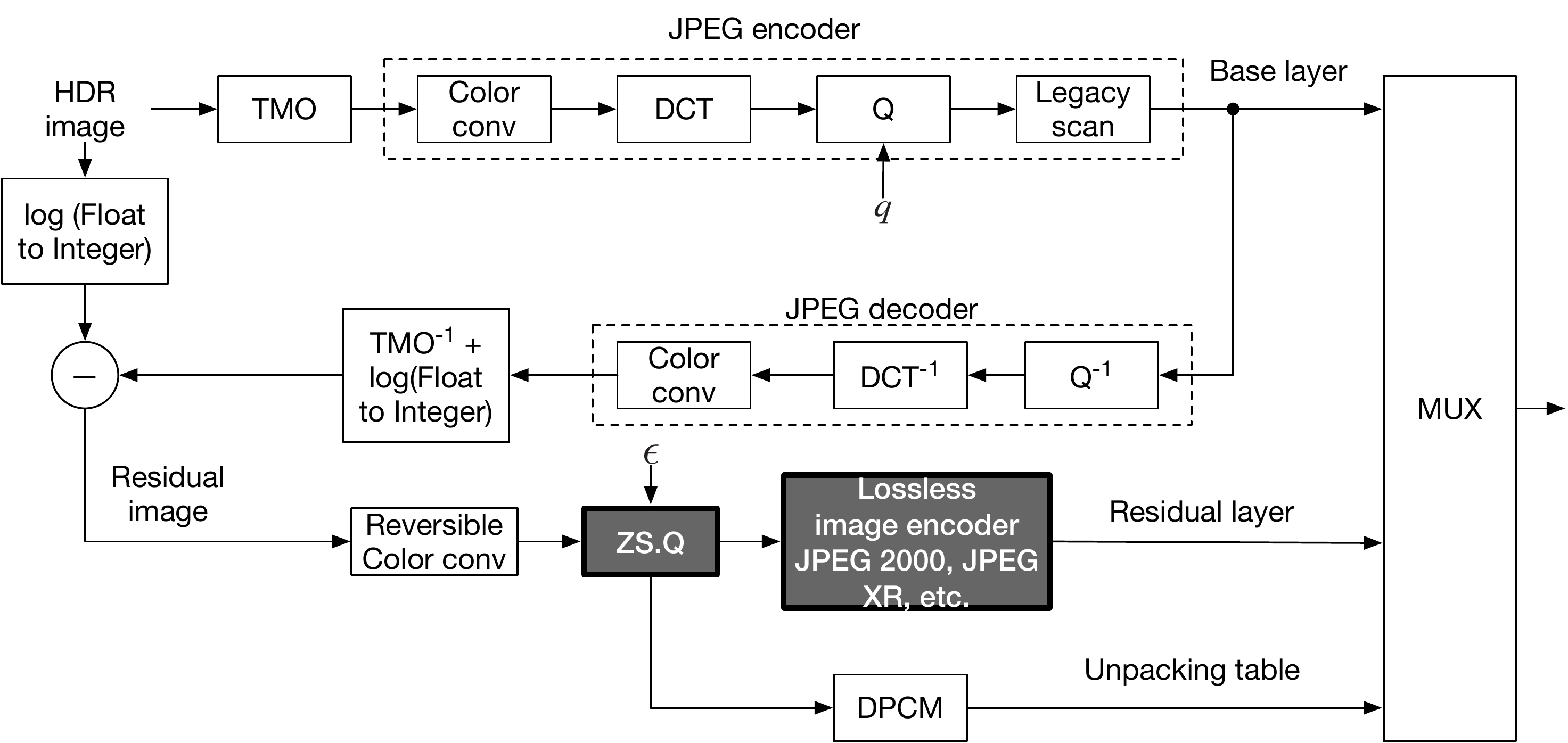}
 \caption{Block diagram of proposed near-lossless two-layer coding method}
 \label{fig:proposed_enc}
\end{figure}
The procedure of the proposed near-lossless two-layer coding method is illustrated in \wfig{proposed_enc}.
The coding-path to generate the base layer, which is backward compatible with the legacy JPEG standard, is the same as that of the JPEG XT Part 8 method except that the refinement scan for the base layer is not used.
Therefore, the LSB split-off in the JPEG XT Part 8 method is not carried out and $R$ is set to zero.

For the residual layer, which consists of the residual data generated
by subtracting the partially decoded base layer from the original HDR image, the coding procedure after color space conversion from RGB to YCbCr differs from the Part 8 method.
The histogram of each color component of the color-converted residual data is quantized using ZS.Q\cite{Minewaki2019}.
The packed residual data are then compressed using an arbitrary lossless image encoder, such as JPEG 2000 or JPEG XR.
Bitrates and image distortion in the near-losssless coder are controlled by a quantization step size parameter $\epsilon$ used in ZS.Q.

For the inverse operation of histogram packing, an unpacking table is
sent to the decoder. The unpacking table is a one-to-one correspondence function between the packed index value and original pixel value.
Since this monotonically increases, Differential Pulse-Code Modulation (DPCM) may be carried out, then the unpacking table encoded using DPCM is compressed by bzip2\cite{bzip2} to reduce the amount of unpacking-table data.

Finally, the base layer which is compatible with the legacy JPEG standard, the extension
layer consists of the lossless JPEG 2000 or JPEG XR codestream, and
the unpacking table compressed using bzip2 are multiplexed into a codestream and sent to the decoder.

\subsection{Zero-Skip Quantization for residual images\label{sec3.1}}
Two-layer lossless coding based on the histogram-packing technique has been shown to outperform JPEG XT Part 8's lossless coding\cite{8351220,8456254}.
However, the histogram-packing technique can not be applied to near-lossless coding, due to not having any quantization.

To extend the histogram-packing technique to two-layer near-lossless coding for HDR images, we focused on ZS.Q\cite{Minewaki2019}.
ZS.Q takes into account not only histogram sparseness but also the effect of quantization under a near-lossless condition.
However, ZS.Q has never been applied to two-layer near-lossless coding.

Let $e$ be a quantization error and, defined by
\begin{equation}
  e = x' - x, \,\,\,\,\forall x\in \mbox{image},
\end{equation}
where $x, x' \in Z$ are a pixel value and dequantized pixel value, respectively.
In near-lossless coding, the quantization errors should be carefully considered to ensure the following equation:
\begin{equation}
|e| \leq \delta,\,\,\,\, \forall\ e,
\end{equation}
and $\delta \in Z$ is a specified value.
That is, $\delta$ means the maximum error between $x$ and $x'$ in an image.

\begin{figure}[tb]
\centering
 \subfloat[Histogram $H(x)$ (before quantization)\label{fig:howtoZSQ:Hx}]{
  \includegraphics[width=0.6\linewidth]{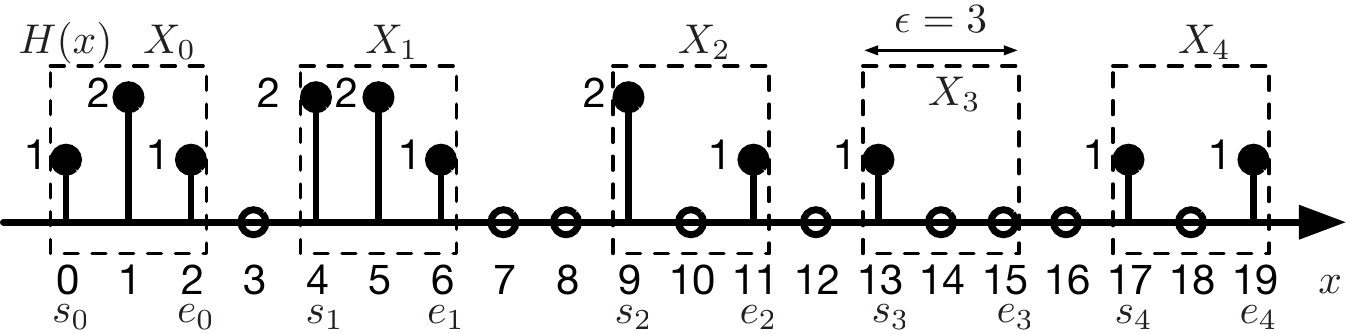}
 }
 \newline\\[-1mm]
 \subfloat[Histogram $H(q)$ (after quantization)\label{fig:howtoZSQ:Hq}]{
  \includegraphics[width=0.6\linewidth]{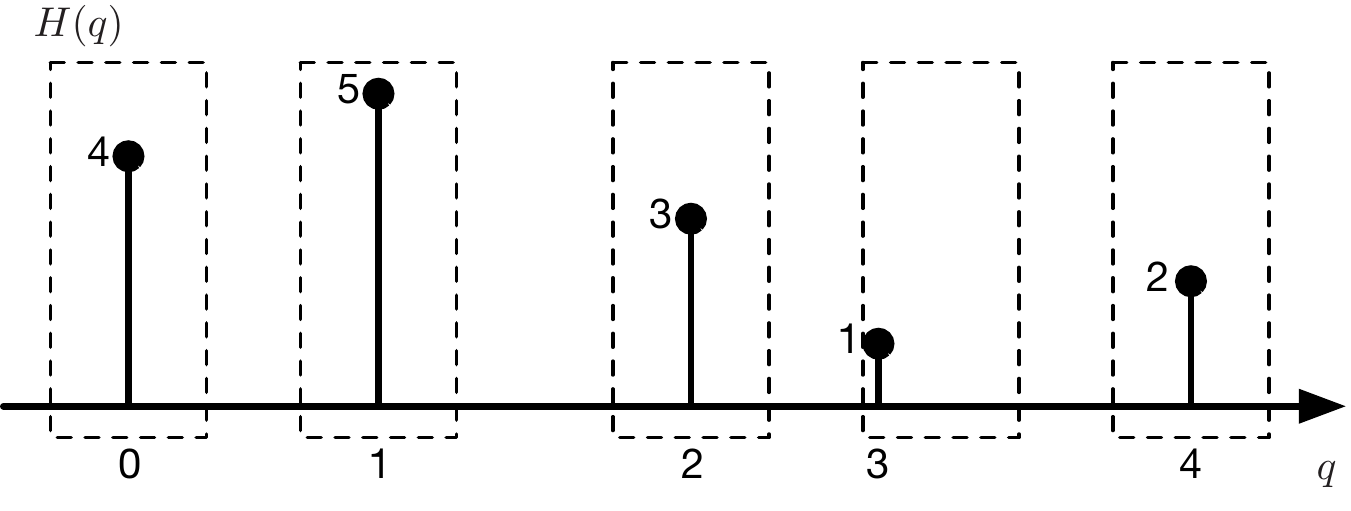}
 }
 \newline\\[-1mm]
 \subfloat[Histogram $H(x')$ (after dequantization)\label{fig:howtoZSQ:Hxd}]{
  \includegraphics[width=0.6\linewidth]{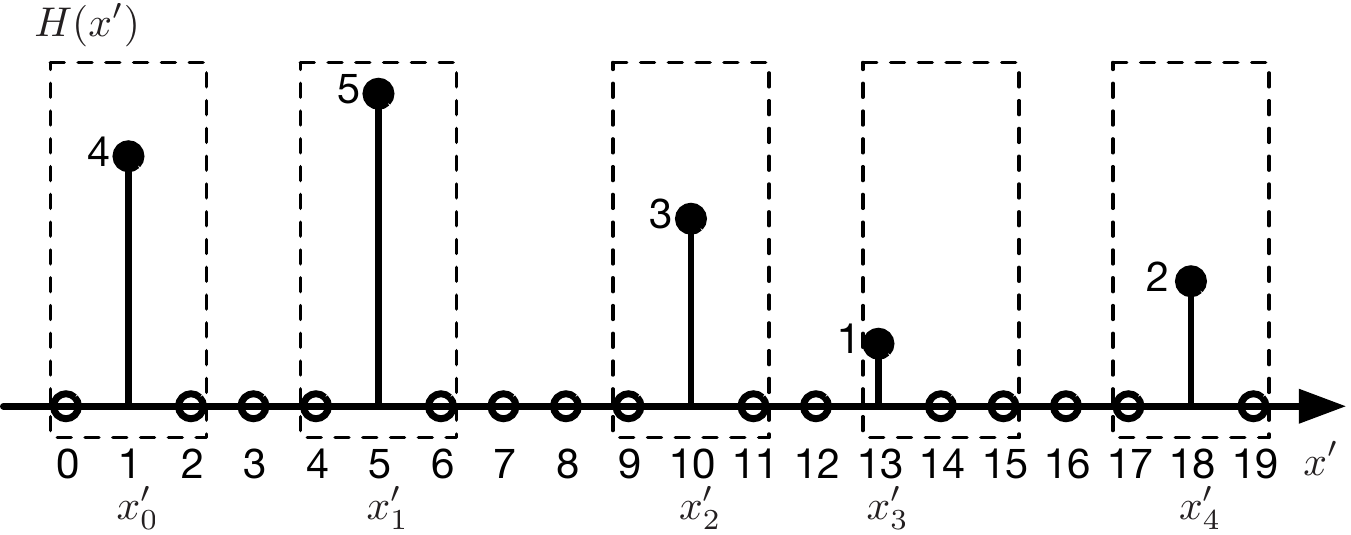}
 }
 \newline\\[-1mm]
 \caption{Example of ZS.Q ($\epsilon = 3$, $\delta = 1$)}
 \label{fig:howtoZSQ}
\end{figure}

The procedure of ZS.Q is illustrated in \wfig{howtoZSQ} and summarized as follows:
\begin{enumerate}
\item Choose a maximum error parameter $\delta$.
\item Choose a quantization step size $\epsilon$ from $\delta$ as
\begin{equation}
 \epsilon = 2\delta\ \ \mbox{or}\ \ 2\delta + 1.
\end{equation}
\item Calculate a histogram $H(x)$ of an input image, as illustrated in \wfigsub{howtoZSQ}{howtoZSQ:Hx}.
\item Classify an input $x$ into a set $X_q$ ($q=0, 1\cdots$) as
\begin{equation}
 X_q = \left\{\,x\,|\,(x\geq s_q)\wedge(x\leq e_q)\,\right\},
\end{equation}
where
\begin{align}
 s_q &= 
 \left\{\begin{array}{l@{}c}
  \min\left\{\,x\,|\,(x\geq 0) \wedge (H(x)\neq 0)\,\right\}, &(q=0) \\
  \min\left\{\,x\,|\,(x>e_{q-1}) \wedge (H(x)\neq 0)\,\right\}, & (q>0)\\
 \end{array}
 \right.\\
 e_q &= s_q + \epsilon - 1
\end{align}
\wfiguresub{howtoZSQ}{howtoZSQ:Hx} illustrates an example of classification of the input value when $\epsilon=3$.
\item Convert all $x$ to the $q$ indexes and generate an index image.
This procedure corresponds to the histogram-packing technique.
\wfiguresub{howtoZSQ}{howtoZSQ:Hq} illustrates histograms $H(q)$ for the index image.
They are denser than that of the original residual image \wfigsub{howtoZSQ}{howtoZSQ:Hx}.
\item Calculate representative values $x'_q$ as
\begin{equation}
 x'_q = \left\lfloor\frac{s_q+t_q}2+0.5\right\rfloor
\end{equation}
where
\begin{equation}
 t_q = \max\{\,x\,|\,(x\in X_q)\wedge(H(x)\neq 0)\,\}.
\end{equation}
In the dequantization process, the $q$ indexes are converted to $x'_q$, referred to as an unpacking table.
As a result, all $x'$ are generated.
\wfiguresub{howtoZSQ}{howtoZSQ:Hxd} illustrates a histogram $H(x')$ for the dequantized image reconstructed by the index image and unpacking table.
\end{enumerate}

\begin{figure}[tb]
   \centering
   \subfloat[original residual data $H(x)$\label{fig:exampleHist:original}]{
     \includegraphics[width=1.0\linewidth]{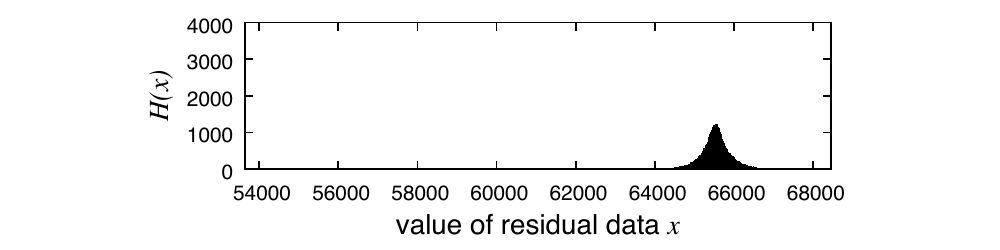}
   }
   \newline\\[-1mm]
   \subfloat[index image  $H(q)$ ($\epsilon=3$)\label{fig:indexTable:index3}]{
    \includegraphics[width=1.0\linewidth]{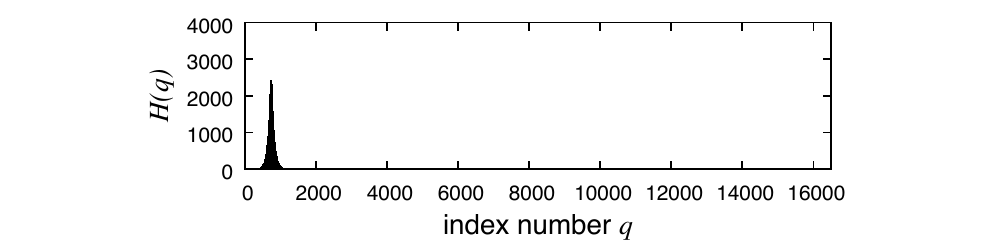}
   }
   \newline\\[-1mm]
   \subfloat[index image  $H(q)$ ($\epsilon = 5$)\label{fig:indexTable:index7}]{
    \includegraphics[width=1.0\linewidth]{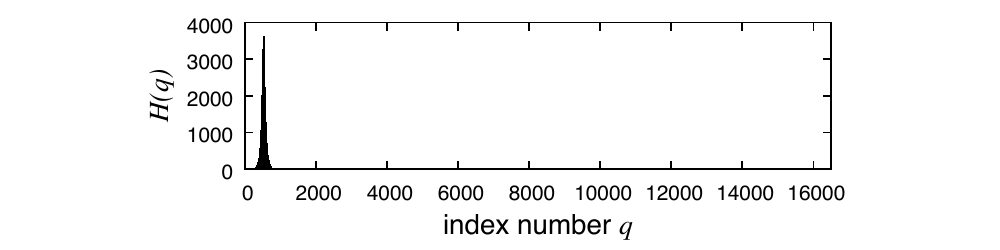}
   }
   \newline\\[-1mm]
   \caption{Histograms of residual data and index images (Y component of `memorial',
     $q=80$)}
   \label{fig:exampleHist}
   \end{figure}

In the above procedure, the quantization error of pixels is controlled by one parameter, $\epsilon$.
\wfiguresub{exampleHist}{exampleHist:original} shows an example of a $H(x)$ for the Y component of the residual data for the HDR image `memorial' under $q=80$.
The horizontal axis denotes $x$, which are mapped to integer numbers from floating-point ones.
Compared with \wfigsub{exampleHist}{exampleHist:original}, \wfigsubonly{exampleHist}{indexTable:index3} and \wfigsubonly{exampleHist}{indexTable:index7} show histograms $H(q)$ for index images produced using ZS.Q under $\epsilon = 3$ and $\epsilon = 5$, respectively.
Due to the increase in $\epsilon$, the range in the histograms becomes narrow, which means high correlation among pixels.

   \begin{figure}[tb]
   \centering
    \includegraphics[width=1.0\linewidth]{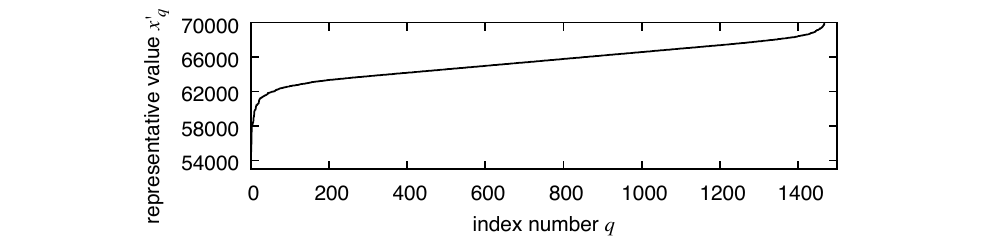}
   \caption{Unpacking table (Y component of `memorial', $q=80$, $\epsilon = 3$)}
   \label{fig:indexTable}
   \end{figure}

An unpacking table, which is necessary to carry out inverse histogram
packing, is illustrated in \wfig{indexTable}. 
This table is monotonic, so it is effectively compressed using DPCM.

\section{Experimental results}
The compression performance of the proposed method was compared with that of the JPEG XT Part 8 method.
For the JPEG XT Part 8 method, the reference
software\cite{JPEGXTpart5} provided by the JPEG committee was used.
For the proposed method, we made modifications to the reference software.
These modifications were to the residual path in accordance with \wfig{proposed_enc}.
The Kakadu software\cite{Kakadu} (JPEG 2000 codec) was also used in the residual path as the lossless image encoder for the proposed method.

\subsection{Rate-distortion curves}
\begin{figure}[tb]
 \centering
 \subfloat[Proposed method and JPEG XT Part 8 method \label{fig:proposedMcKeesPub}]{
  \includegraphics[width=0.7\linewidth]{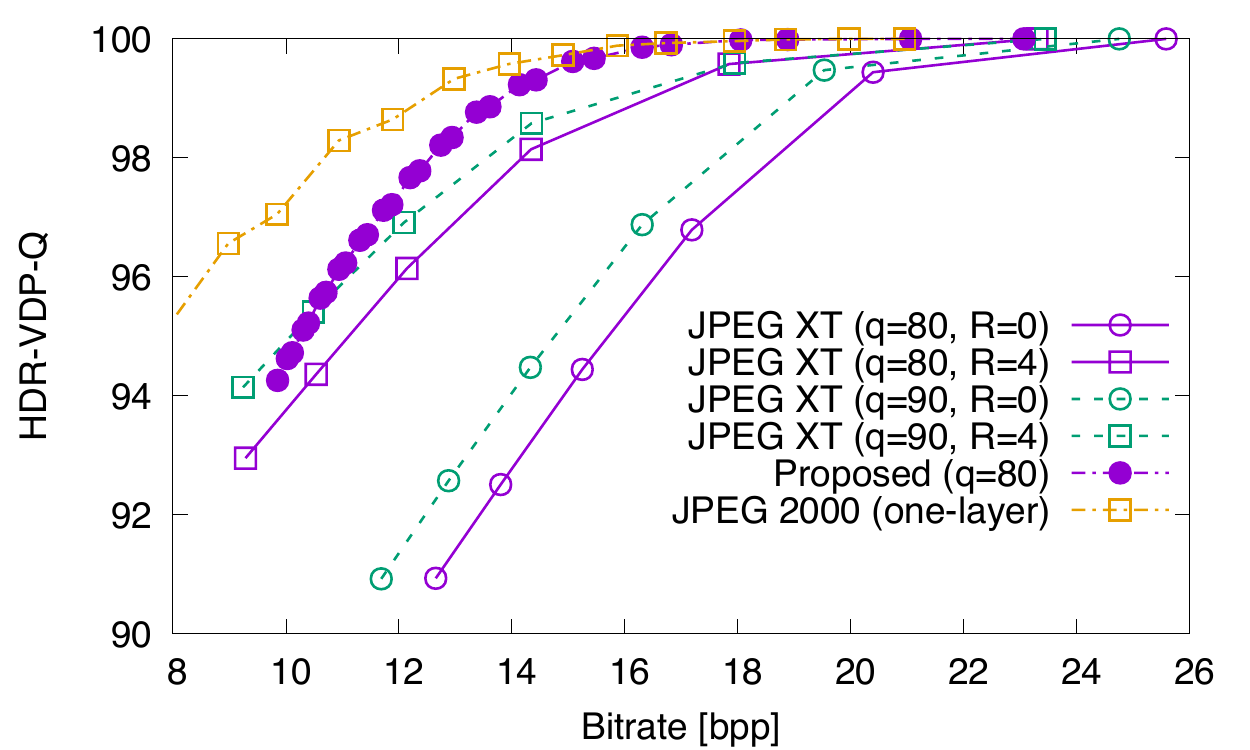}
  }
  \newline\\[-1mm]
 \subfloat[Proposed method\label{fig:proposedOnly}]{
  \includegraphics[width=0.7\linewidth]{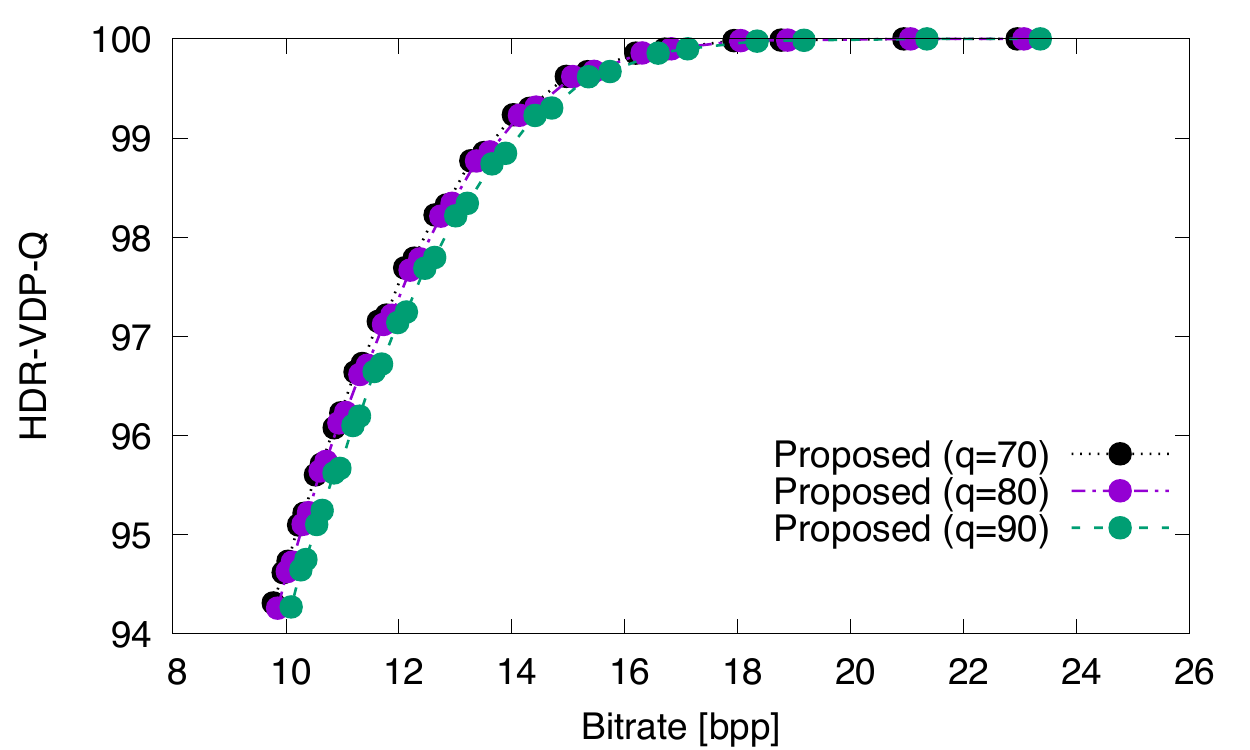}
  }
 \caption{Rate distortion curves of proposed method and JPEG XT Part 8 method (McKeesPub with 48bpp)}
 \label{fig:nearLosslessProposed}
\end{figure}

\wfiguresub{nearLosslessProposed}{proposedMcKeesPub} shows the comparison of compression performance for an HDR image,
McKeesPub, between the proposed method and the JPEG XT Part 8 method.
As a reference for one-layer, the results of the JPEG 2000 coding method with reversible color transform, reversible Discrete Wavelet Transform (DWT) and without quantization (only bit-truncation was used) is also illustrated.
The bitrate was controlled by $Q$ for the JPEG XT Part 8 method under fixed two parameters $q$ and $R$, and by $\epsilon$ for the proposed method.
In \wfigsub{nearLosslessProposed}{proposedMcKeesPub}, there are six control points from $Q=95$ to $100$ for the JPEG XT Part 8.
In contrast, there are twenty-nine control points, from $\epsilon = 1$ to $29$.
The proposed method provided much finer rate control than the JPEG XT Part 8 method.

\wfiguresub{nearLosslessProposed}{proposedMcKeesPub} clearly shows that the
proposed method outperformed the JPEG XT Part 8 method in terms of HDR-VDP-Q.
The HDR-VDP-Q values of the proposed method were maintained at more than 98, even when choosing 13bpp as a target bitrate.
The proposed method also had almost the same performance as JPEG 2000 coding  above 16bpp, regardless of using a two-layer structure. 

\wfiguresub{nearLosslessProposed}{proposedOnly} shows  rate-distortion curves of the proposed method with various $q$ values.
The proposed method also has almost no effect to HDR-VDP-Q and bitrates, although
$q$ in the JPEG XT Part 8 method had a much stronger impact on the HDR-VDP-Q than that of the proposed method. 
In other words, the compression performance of the proposed method depends on only one parameter, $\epsilon$, although the JPEG XT Part 8 has three parameters, $R$, $q$, and $Q$.

 \begin{figure}[t]
  \centering
    \subfloat[Q=99\label{fig:compareQ99}]{
     \includegraphics[width=0.3\linewidth]{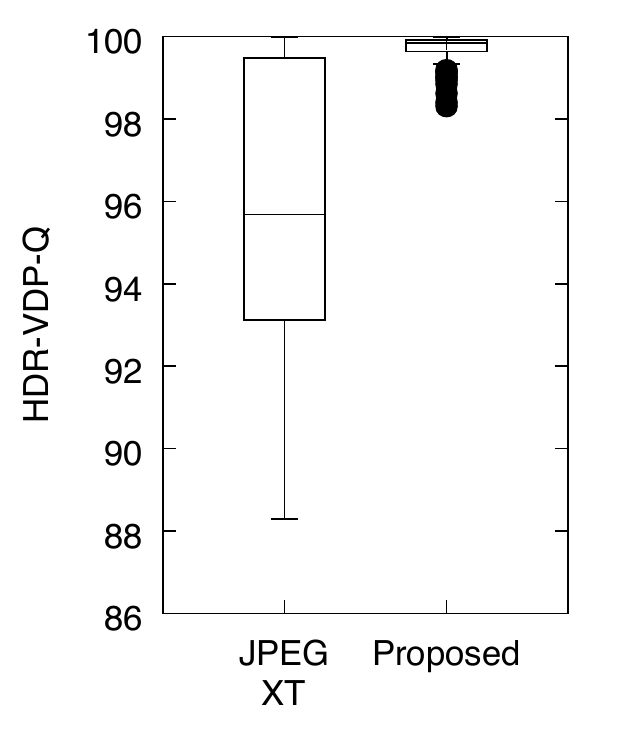}
    }
    \subfloat[Q=98\label{fig:compareQ98}]{
     \includegraphics[width=0.3\linewidth]{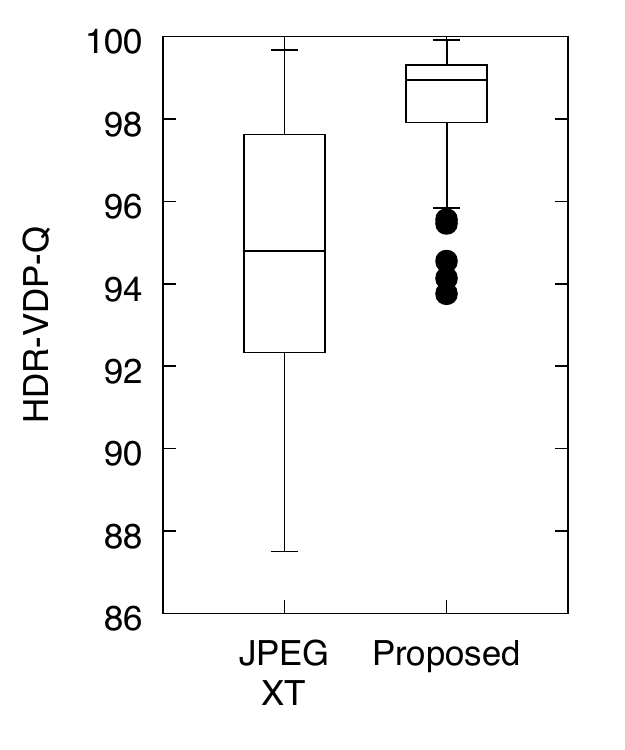}
    }
    \subfloat[Q=97\label{fig:compareQ97}]{
     \includegraphics[width=0.3\linewidth]{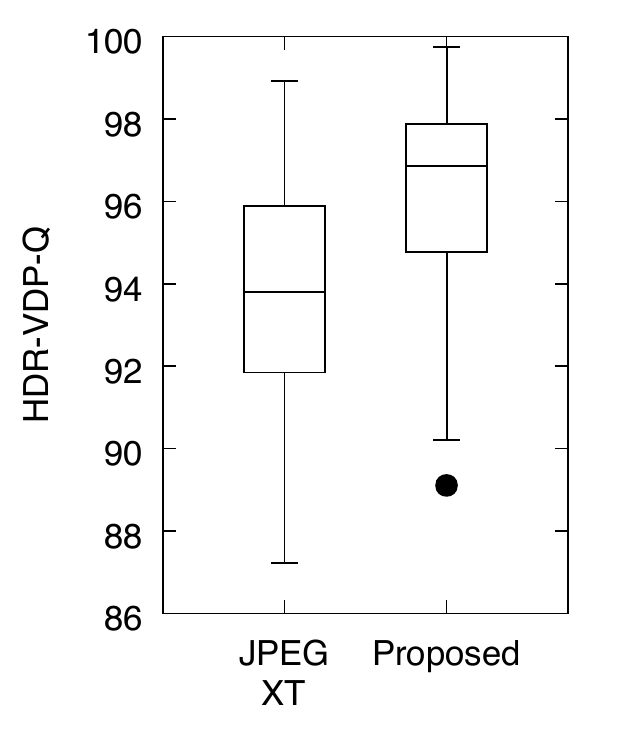}
    }
  \caption{Boxplot of HDR-VDP-Q of proposed method ($q=80$) and JPEG XT Part 8 method ($q=80, R=4$).
  Box indicates three quartiles of data, and whiskers indicates lowest datum within 1.5 interquartile range (IQR) of lower quartile, and highest datum still within 1.5 IQR of upper quartile.
Data not included between whiskers should be plotted as outlier with dot.}
  \label{fig:compBitrate}
 \end{figure}

\subsection{Comparison of compression performance using 105 images}
\wfigure{compBitrate} shows boxplots of HDR-VDP-Q calculated using 105 HDR images collected from the Fairchild HDR image survey \cite{fairchild2007hdr}.
For the JPEG XT Part 8 method, we chose $R=4$, $q=80$, $Q=97, 98$, and $99$, so  $105\times 3$ JPEG XT bitstreams were generated in total.
For the proposed method, the bitrates were set to almost the same as those of the JPEG XT Part 8 method under each condition.
From \wfig{compBitrate}, the proposed method outperformed the JPEG XT Part 8 method.
%

\section{Conclusions}
We proposed a novel two-layer near-lossless coding method using zero-skip quantization (ZS.Q) for HDR images with backward compatibility with the legacy JPEG standard.
The proposed method was demonstrated to solve two issues that the JPEG XT Part 8 method has: near-lossless compression performance and combination of coding parameters.
While compression performance of the JPEG XT Part 8 was controlled by three parameters: $q$, $Q$ and $R$, that of the proposed method was controlled by only one parameter: $\epsilon$.
The experimental results also shows that the proposed method has a higher compression performance than the JPEG XT Part 8 method in terms of HDR-VDP-Q.

\bibliographystyle{IEEEbib}
\bibliography{./bibs/IEEEabrv,./bibs/refs}

\end{document}